\documentclass[aps,amssymb,amsmath,amsfonts,twocolumn]{revtex4}
\usepackage[dvips]{graphicx}
\usepackage{bm,bbm}

\usepackage[usenames]{color}

\begin{document}
\title{Citation graph, weighted impact factors and performance indices}

\author{Karol {\.Z}yczkowski$^{1,2}$}
\affiliation{$^1$``Mark Kac'' Complex Systems Research Centre, Institute of Physics,
   Jagiellonian University, ul. Reymonta 4, 30-059 Krak{\'o}w, Poland}
\affiliation{$^2$Centrum Fizyki Teoretycznej, Polska Akademia Nauk,  
Al. Lotnik{\'o}w 32/44, 02-668 Warszawa, Poland}

 \date{March 11, 2010}

\begin{abstract}
A scheme of evaluating an impact of a given scientific paper
based on importance of papers quoting it is investigated.
Introducing a weight of a given citation, dependent on the 
previous scientific achievements of the author of the citing paper,
we define the weighting factor of a given scientist. Technically the weighting 
factors are defined by the components of the normalized leading eigenvector 
of the matrix describing the citation graph.
The weighting factor of a given scientist,  
reflecting the scientific output  of other researchers quoting his work,
allows us to define weighted number of citation of a
given paper, weighted impact factor of a journal and  weighted
Hirsch index of an individual scientist or of an entire scientific institution.
\end{abstract}

\maketitle

\medskip

\section{Introduction}

Any given scientist is a good scientist if he is considered
to be good by a representative group of other good scientists.
Such a simple way of evaluating quality of scientific achievements 
could be useful two hundred years ago, as the number of 
scientists was small and a respectable researcher 
was competent to evaluate the progress in a huge field of science.

Nowadays such an approach is no longer realistic.
As the number of universities, scientists, journals and 
scientific articles keeps growing fast,
one is often forced to use some parametric measures 
to characterize the output of a given scientist. Although 
the peer review is  still considered to be the most reliable and objective
method of evaluation of candidates for any scientific position,
in view of a large number of applicants
in the preliminary phase one often performs 
screening of numerical values of performance indices, 
designed to quantify scientific output of the candidates.

As the status of scientific citations among researchers is rather ambivalent
\cite{AR09}, we do not claim that 
citations of scientific articles directly indicate their quality and importance. 
Just on contrary, we share some doubts, often raised  in the literature
\cite{AET08,LM09} that
trying to measure scientific achievements by any index based on
the number of citations may generate certain perverse effects:
researchers no longer focus on interesting and relevant research,
but they try to adapt to the popular evaluation criteria.
However, looking around  we have to agree that various citation indices
are used nowdays to evaluate scientists, journals or research institutions.

Thus in this work we shall not discuss a controversial issue, 
what is the optimal measure of scientific achievement.
Instead we review common quantitative measures
of scientific quality and discuss possible ways to improve them.
The most popular indices used to evaluate 
the impact of a given article, the influence
of a scientific journal for the research community,
the scientific output of a single individual or entire institution
are based only on  the quantity of citations in the literature
to the articles analyzed. Our aim is to take  into account
also the quality of the citations,
measured by the averaged achievements of the authors
of the papers which refer to the article under consideration.

Additional motivation for our research is due to
the controversy concerning the usage of the impact factor (IF) \cite{gar1,gar2},
to quantify the quality of a scientific journal.
On one hand it was pointed out \cite{FG08}
that impact factor of a given journal can be manipulated
by the editors and publisher.
On the other hand it was often emphasized
that the two year window for counting the citations of the papers analyzed
is perhaps fine for biology, medicine and some other branches of science, 
it is rather not the case e.g. for mathematical journals.
These journals score small values of the impact factor, since 
the preparation of a mathematical article and the entire refereeing procedure
takes often more time than two years.
Furthermore, in several branches of science 
great role is played by articles which are not quickly forgotten.
Thus one could also design and work with an impact factor,
which takes into account only citations gathered 
three or five years after paper was published \cite{LM09}.

In order to identify the papers which contribute most to 
the IF of a journal the editors often try to identify articles,
which gained the largest number of citations in the first and 
the second year after the year they were published.
Editors of some mathematically oriented journals,
analyzing a list of articles prepared in this way for their journal
were concerned, if it really represents the most important articles published.
In fact they considered that several papers of not the top quality
entered this list, only because they were simple enough
that they could become understood and later quoted 
by other authors of recent papers of a mediocre quality.

On the other hand, in is also believed that in several 
fields of science the impact factor could be
artificially inflated by a number of papers of lesser quality,
the authors of which tend to quote several recent articles
not directly related to their work.
The aim of this practice is to please the editor
(if the paper cited was published in the same journal),
or to suggest the referees that the author 
follows the recent literature and in this way
to improve the chances that their work will be published.

To take these features  into account one needs then 
to distinguish, in a statistical sense, the quality 
of a given citation.
In short, any citation of an established scientist, whose numerous 
papers have already attracted several citations, 
should be weighted more than a citation by a newcomer to the field.
In this paper we suggest a possible
solution of this problem by defining the  weight of a scientific citation
and using this notion to modify and  improve  performance  indices 
defined  earlier  in \cite{gar2,Pl94,Hi05}.

All indices proposed are based on the weighting factor, associated to each scientists,
which is analogous to the Page Rank introduced by Brin and Page \cite{BP98}
to characterize relative importance of various web pages
and used in the Google web search engine.
The weights defined by the components of the leading eigenvector
of a suitably defined citation matrix  characterize a given citation.
These numbers  display the desired property of  self-consistence: 
the weight of any citation by a given researcher is larger, 
if his papers are quoted by other scientists, whose papers are often quoted.

A similar idea was recently applied for study of the citation graph 
created for publications in the Physical Review family of journals \cite{CXMR07},
for the graph in the field of
biochemistry and molecular biology \cite{MGZ08},
and independently  put forward in recent lectures of Nielsen \cite{Ni08},
who considered the idea to use the Page Rank algorithm to order
individual scientific papers according to their citation graphs.
The same algorithm was used to design the {\sl Eigenfactor}
web tool, which takes into account the citation graph to evaluate a
proposed measure of the relative importance of scientific journals \cite{Be07}.
While any node of the graph represents a single article
in the former approach or an entire journal in the latter scheme,
in this work it will be associated with an individual scientist.

\section{Citation matrix and weighting factors}

Consider a sample of $N$ authors of numerous scientific articles, 
in which they usually refer to their previous achievements, 
but also quote papers of some other scientists.
Let us assume for a while that all papers considered are written by a single author only -
this simplifying assumption will be relaxed later in this section.
Each scientist can be thus associated with a vertex of a graph,
while any citation in any paper forms a directed link between two vertexes -- 
see e.g. \cite{AJM04}. 
Define a square matrix $C$ of size $N$, such that 

a) $C_{ij}$ is equal to the number of times the scientist "$j$" quoted a 
        single paper of his colleague $"i"$.

b) $C_{ii}=0$, for $i=1,\dots,N$, hence all self-citations are neglected.

Observe that this {\sl citation matrix} $C$ is 
likely not to be symmetric. 
However,  matrix $C$ is by construction real and it contains
non-negative entries only. 
 Therefore it fulfills the assumptions of the celebrated 
{\sl Frobenius--Perron  (FP) theorem}  
(see  e.g. \cite{MO79,Ber05:1} ). This implies that

i) there exists an eigenvalue $z_1=\lambda$ with the largest absolute
 value which is real and non-negative,

ii) the entire spectrum $\{ z_i\}_{i=1}^N$ of $C$ belongs to the 
disk of radius equal to $\lambda$,

iii) the eigenspace associated with $\lambda$ contains a
real eigenstate ${\vec x}=\{x_1,\dots x_N \}$, 
such that all its components are non-negative.

\smallskip

\noindent
Since we do not force citation matrix $C$ to be stochastic, 
the leading eigenvalue $\lambda$ needs not to be equal to unity.
However, we will assume here that the graph analyzed is connected.
Then the leading eigenvalue $\lambda$ is non-degenerate,
and  there exists a unique vector  ${\vec x}$ such that
\begin{equation}
  C x = \lambda x . 
\label{eigen}
\end{equation}
This leading eigenvector can be normalized as 
\begin{equation}
W_i \ := N  \frac{x_i}{ \sum_{j=1}^N x_j} ,
\label{normal}
\end{equation}
which implies that the mean entry is equal to unity, $\langle W_i \rangle_i = 1$.

In this way for a given scientist $i$ one can associate an weighting factor $W_i$.
Such a factor depends not only  on total number  of times his papers were quoted by other scientists,
$T_i:=\sum_{j=1}^N C_{ij}$, but also on the fact {\it who} referred to his work.
However, such a number should not be treated as an optimal number used to quantify the
scientific achievement of a researcher. It is more informative then the bare
number $T_i$ of total citations, but it shares similar disadvantages.
For instance, as emphasized by Hirsch \cite{Hi05}, the total number
of citations can be inflated by single non representative papers,
and it overweights highly quoted review articles
versus original research papers. On the other hand
the weighting factors $W_i$ allow us to define other
more suitable indices and parameters.
Before proceeding we need to adjust the definition
of $C$ and its eigenvector $\vec x$,
to make it directly applicable to the problem.

\subsection{Scientific papers with several authors} 
 
Any citation to a single author paper published in an article written by 
another single author can be interpreted as a unit flow between 
the corresponding two vertexes.
Thus the number
\begin{equation}
 L=\sum_{ij} C_{ij}
\label{susum}
\end{equation}
represents the number of unit links in the graph,
equal to the total number of citations 
(with auto citations excluded).

In practice, the papers are often written by several authors, 
so it is natural to split this coupling uniformly 
among all the authors involved in such a way
that the sum of the weights for each citation is equal to unity.
To this end, consider the process of forming a graph
by taking into account one article after another,
and in each case scanning through all its references.

Assume that a paper with $M$ authors defined by the set of indices,
$J=\{j_1,\dots, j_M\}$ quotes another paper by
$K$ authors,  described by the set $I=\{i_1,\dots, i_K \}$.
Any quotation contributes to the citation matrix according to 
two rules:

a') if $I \cap J = \{\emptyset\}$ then  $C_{ij} \to C_{ij}+\frac {1}{KM}$
     for all pairs of indices $i,j$ such that $i\in I$ and $j\in J$.
    In words, an independent citation is taken into account and normalized in 
    such a way that the number $L$ defined by (\ref{susum}) grows by one.

b') if $I \cap J\ne \{\emptyset\}$ then  $C\to C$;
   the citation matrix does not change since one does not want
   to analyze {\sl dependent citations},

 Observe that in the case of all papers written by single author 
the rules a') and b')  reduce
 to the rules a) and b) discussed before.

Let us emphasize that the assumptions a) and a') 
imply that all quantities considered further 
do  not depend on self citations and dependent citations,
which are known to influence bibliometric
indices \cite{Ak03,GDTS06,Sch07,Fr07,FA07}.
Alternatively, one could neglect all auto citations but 
take into account the dependent citations
and attribute to them the weight smaller than this characterizing independent citations.

\subsection{Truncated citation graph}

In practice it is hardly doable to take into account {\it all}
the scientist into the consideration. Even if one could aim to
make the graph as complete as possible, its truncation at some stage seems inevitable.
In any realistic case the citation graph will describe a finite set of $N$ 
researchers and take into account a given number of their papers and the cumulative
list of all the references. This list of citations will likely include 
references to the papers written by authors not belonging to the analyzed set of scientists.
To take into account the fact that papers written by the researcher represented by $i$-th 
vertex of the graph are cited by authors outside the graph  
we suggest to extend the citation matrix by an extra row and extra column,
which jointly represents all truncated vertexes.  
The additional entries read,

c) $C_{i,N+1}$ is equal to the total number of times, the papers of scientist "$i$" were
     quoted by all authors outside the graph (not belonging to the analyzed set),

d) $C_{N+1,i} :=\frac{1}{N} \sum_{k=1}^N C_{ki}$ for $i=1,\dots, N$. 
This assumption is made to 
     attribute a well balanced, average weight to all external citations. 
To be consistent with the rule b) we will also set 

e) $C_{N+1,N+1}=0$.

Note that the last, fictitious, vertex of the graph has no direct meaning,
since it only represents the world outside the graph.
The eigenvector $x$ of the augmented matrix has $N+1$ components,
but  only first $N$ of them have the meaning of the 
weighting factors for $N$ individuals. 
Therefore its last component $x_{N+1}$ can be neglected, and in the normalization scheme 
which eventually produces the vector $\vec W$ of size $N$
the same rescaling (\ref{normal}) can be used.

\section{Weighted performance indices}

\subsection{Weighted number of citations} 

After constructing the complete citation matrix $C$
or its approximation obtained 
according to the rules specified above,
we find its normalized leading vector 
and normalize it as in (\ref{normal})
to obtain the vector of the weighting factors $W_i, i=1,\dots, N$.
Spectra and leading eigenvectors of some exemplary graph matrices 
are discussed in Appendix A, while 
the issue of uniqueness of the vector corresponding to the 
leading eigenvalue is discussed in Appendix B.
For any scientific article $A$ one may find in an appropriate date base
the number of times it was quoted by all other scientific papers in the literature.
Denoting this number by $c(A)$ we are now in position to 
define the weighted number of quotations,
\begin{equation}
w(A) :=  \sum_{j=1}^{c(A)} W_j ,
\label{quot1}
\end{equation}
where $W_j$ denotes the weighting factor of the $j$-th author quoting the paper $A$.
For consistency we are not going to include into this sum any auto citations.
In a more general case of papers written by several authors it is natural
to take the average weight of these authors. Therefore we write
\begin{equation}
w(A) :=  \sum_{j=1}^{c(A)}   \; \frac{1}{n_j} \sum_{\mu=1}^{n_j} W_{j\mu} \ ,
\label{quot2}
\end{equation}
where $c(A)$ represents the number of independent quotations of the paper $A$,
while $n_j$ denotes the number of authors of the $j$-th paper quoting $A$,
and $W_{j\mu}$ is the weighting index of the $\mu$-th co--author of this paper. 

\subsection{Weighted impact factor of a journal} 

Let $Z_y$ denotes the number of papers published by a certain scientific journal $J$
in year $y$.
To quantify the impact the journal exerts for  the scientific community
one often uses the so called {\sl impact factor}. 
To compute it one takes all
 $Z_{y-2}+Z_{y-1}$ articles published one or two years earlier,
 and then sums the number of citations $c(A_j)$,
a given article from this set received during the year $y$. 
 The result has then to be normalized 
 with respect to the total number of articles published in journal $J$
during the two year time span \cite{gar1,gar2},
\begin{equation}
{\rm IF2}_y(J) \  :=  \  \frac{1}{Z_{y-2}+Z_{y-1}}   
             \sum_{j=1}^{Z_{y-2}+Z_{y-1}}  c(A_j)  \  .
\label{IF2}
\end{equation} 
This commonly used index takes into account 
the two year time window, so we shall denote it by IF2.

By construction this quantity takes into account only the 
{\sl quantity} of the citations received by articles published in a given journal
during last two years but not their {\sl quality}.
Presented approach allows us to take into consideration
the fact, {\sl who} quoted the papers analyzed.
In full analogy to (\ref{IF2}) we thus define the
{\sl weighted impact factor} (WIF2) of a journal $J$,
\begin{equation}
{\rm WIF2}_y(J) \  :=  \  \frac{1}{Z_{y-2}+Z_{y-1}}   
             \sum_{j=1}^{Z_{y-2}+Z_{y-1}}  w(A_j) \  .
\label{WIF2}
\end{equation}
The only difference is that instead of counting the bare numbers of citation $c(A_j)$
of a given article $A_j$,  
we now sum the weighted citations  $w(A_j)$.
Since these number reflects in a sense the quality of a citation
we tend to believe that the weighted impact factor forms a more
accurate quantity to evaluate the quality of a
scientific journal than  the standard IF.

As mentioned in the introduction in some disciplines
like mathematics and mathematical physics the 
process of preparing an article and publishing it is often longer than the two year time
span used in the definition of IF2. Therefore one may propose \cite{AET08}
to use also similar quantities defined for a longer time window containing 
{\sl five} years. In full analogy to the previous definitions we write
\begin{equation}
{\rm IF5}_y(J)  \ :=   \  \frac{1}{m}   
               \sum_{j=1}^{m}  c(A_j)  
{\rm \quad where \quad}
m=\sum_{i=1}^{5} Z_{y-i} 
\label{IF5}
\end{equation} 
and
\begin{equation}
{\rm WIF5}_y(J) \  :=  \  \frac{1}{m}   
             \sum_{j=1}^{m}  w(A_j)  
 \ .
\label{WIF5}
\end{equation} 

Here one takes into account all articles 
published in the five year time span, while 
$c(A_j)$ and $w(A_j)$ denote now the number of all citations
and the sum of weighted citations, a given paper $A_j$ from this
sample obtained in the analyzed year $y$.
The $5$--year impact factor could be specially useful  
to characterize  mathematical journals 
and journals devoted to these fields of science,
in which the papers are produced in a slower pace,
and the citations half--life is longer, since
after a few years the articles do not become obsolete. 

\subsection{Weighted impact factor of a paper} 

Since the distribution of citations is known to be skew 
      \cite{Se97,AET08,RFC08}
the providing the average number of citations only 
is by far not sufficient to characterize the entire distribution.
Hence it is not possible to use the impact
 factor of a journal as an estimated.
 number of the citation a typical paper published there 
   will obtain during the next two years.
Moreover, es explicitly emphasized by Seglen \cite{Se97},
the number of citation obtained by a given article 
is not influenced by the impact factor of the journal,
in which it appeared.

To make any reasonable evaluation of the impact
a given article had on the scientific community,
one can  analyze its contribution to the impact factor of the journal.
To this end we define the {\sl impact factor of an article} $A$
published in year $y$ is a sum of citations gained in 
the next two years,
\begin{equation}
{\rm AIF2}_y(A) \  :=  \  c_{y+1}(A) +c_{y+2}(A) \  ,
\label{AIF2}
\end{equation} 
since only this citations contribute to the impact factor IF2.
Here $c_y(A)$ denotes the number of times the paper $A$ was quoted 
during year $y$.
Note that the article impact factor (AIF2)
can be only defined only for articles published more than two years ago.
In view of the statistical properties of the citation distribution
it is  clear therefore that for any paper older then two years 
this very quantity has to be used to describe its impact on the field,
instead of the IF of the journal it was published.
In a similar manner, for papers older than five years
one can also define the five years impact factor (AIF5).

To take also into account the quality of each citation
we can use of the weights $w_y$ introduced in the previous section
and define the {\sl weighted article impact factor} (WAIF2) 
\begin{equation}
{\rm WAIF2}_y(A) \  :=  \  w_{y+1}(A) +w_{y+2}(A) \  ,
\label{WAIF2}
\end{equation}
where $w_y(A)$ denotes the sum of the weighted citations 
the paper $A$ defined as in (\ref{quot2})
for citations gained during the year $y$.
By construction this notion is applicable to
articles published at least two years earlier.

\subsection{Weighted Hirsch index} 

To quantify a scientific research output of a given researcher
one often uses the $h$ index introduced by Hirsch \cite{Hi05}.
For a given scientists this index is equal to $h$,
if $h$  of all papers he has written were quoted at least $h$ times.
Ordering his articles according to the number
of citations $c(A_i)$, the article $A_i$ has ever received
one can write
\begin{equation}
h   := \max k : c(A_k) \ge k  \ .
\label{hin}
\end{equation} 

Although the $h$ index gained considerable popularity
and it became a subject of several research papers 
\cite{Gl06,Hi07,Sch07,Wo08},
several of its drawbacks were emphasized \cite{AET08,LM09}.
As in the case of the impact factor the Hirsch index is not capable 
to differentiate between relevant and less relevant citations.

Making use of the weighted citation number $w(A_i)$ of an article,
we may thus introduce the
{\sl weighted $h$ index} 
\begin{equation}
w   :=  w(A_k) 
\label{whin}
\end{equation} 
where $k$ is the maximal integer such that  $w(A_k) \ge k$.
This index enjoys all the virtues of the original $h$
index recently  emphasized in \cite{Hi07,Gl06},
but additionally it takes into account scientific achievements 
of the authors quoting the work of the evaluated individual.
Since the weights $w(A_i)$ determined by the
citation graph  take into account the
number of authors of the paper, the weighted index $w$
does not suffer a crucial drawback \cite{LM09} of the original
Hirsch index, in which a paper with a hundred co--authors is treated in the same way
as an article written by a single scientist.
Furthermore, the weights $w$ are in general given by real numbers,
so the index $w$ may admit non-integer numbers. Thus
this quantity provides us a finer differentiation of the group analyzed
than the index $h$, which is integer by definition.

A possibility to use the  $h$-index 
to quantify scientific production of an entire institution
was recently advocated in  \cite{MM08}.
Hence one can use the weighted index $w$ for this purpose.
Furthermore, following Schubert \cite{Sc07}
one can easily adopt his idea of successive performance indices 
and define an analogue to the index $h_2$.
To be concrete, the {\sl weighted successive index}  $w_2$
of a scientific institution is equal to an integer number $w$,
if it employs $w$ scientists, such that the
weighted Hirsch index $w^{(i)}$ for each of them is equal to or larger than  $w_2$.

\subsection{Weighted efficiency index} 

Let us emphasize here that one should not directly 
compare the Hirsch indices for scientists working in 
different research fields. This is due to the fact that the 
numbers of papers and citations vary from one 
scientific field to another \cite{AET08,BCKM06,AWBB09},
so the means values of Hirsch indices also do differ.
On one hand one could compare the
values of the indices rescaled against the average value
in a given field \cite{RFC08}. On the other hand,
one may work with other indicators which reflect
citing patterns of each community.

As an example of such a quantity one 
consider the number of 'known papers'
produced by a given researcher.
Defining the {\sl known paper}
as an article quoted more times then the number
of references cited in it, we see that
this notion by construction
takes into account the citation habits of a given field.

To set a simple normalization scale useful for comparison
of citations gained by articles from various disciplines
Plomp \cite{Pl94} introduced the {\sl efficiency} of a given paper.
It is defined by a ratio,
\begin{equation}
E(A) = \frac{c(A)}{r(A)} \ ,
\label{eff}
\end{equation} 
 where $c(A)$ denotes the number of citations gained,
while $r(A)$ is equal to the number of articles quoted in work $A$.
In such a way various citation habits, different in different fields
of science are automatically taken into account.  
Moreover, the role of a single citation of a review paper seems to be adequate,
since a good review may attract a lot of citations, but its list of references
is usually also long.

In a loose analogy to the Hirsch index,
we can define an {\sl efficiency index}, ($e$-index), 
which quantifies the research output of a given researcher,
\begin{equation}
e  :=   \max k : E(A_k) \ge 1  \ .
\label{eff1}
\end{equation} 
In other words, for a given author we count the number of his scientific 
papers, which belong to the class of  'known papers' - 
they have gained more citations, 
than the number of items in the list of references in this article.

In spirit of this work we may improve this quantity and define the
{\sl weighted efficiency index} 
\begin{equation}
e'  :=   \max k : w(A_k) \ge r(A_k)  \ .
\label{eff2}
\end{equation} 
Now we count the number of articles
for which the weighted number of citations $w$ is larger or equal to the
total number of references $r$. Such an index is perhaps not as sophisticated 
as the Hirsch index, but its values are by construction less dependent 
on the working habits in a given scientific discipline.

\section{Concluding Remarks}

Analyzing the entire citation graph and citation matrix
one can obtain a weighting factors which quantify the  
total impact of a single researcher for the scientific literature.
It will be interesting to analyze statistical distribution
of weighting factors for the citation graph
representing the entire scientific literature and certain particular branches 
of science.  An  empirical study of papers on high energy physics 
 \cite{Re98,LLJ03} and  computer science \cite{AJM04}
reveals that the probability $P(k)$
 that a given article is cited $k$ times, decays according to a power law,
 $P(k)\sim k^{-a}$. 
A power law distribution of the weighting factors attributed to individual scientific papers on 
molecular biology and  biochemistry  was recently reported by Ma et al. \cite{MGZ08}.
Thus one could verify, whether a similar behavior will be observed for the distribution of 
weighting factors characterizing the group of scientists working in a given field.

The weighting factors attributed to a given scientist
are useful to introduce further bibliometric quantities.
For a given article one defines its weighted number of citations,
for a journal its weighted impact factor
and for a given scientists the $w$ index, i.e. his weighted Hirsch index.

Analogous quantities can be introduced
for groups of researchers or entire scientific institutions,
but their normalization and interpretation has to be
performed with  a certain caution \cite{MM08}.
Similarly, cumulative $w$ indices can be used for
various scientific fields and sub-fields 
just to identify so called 'hot topics' \cite{Ba06}.
The usage of the weighted indices in all these cases
could be superior with respect to the standard quantities 
in a sense that the approach proposed takes into account 
the average quality of the citations of scientific articles.

However, it should be emphasized explicitly 
that the computation of weighted scientometric indices  
is it not entirely straightforward and for
any practical purpose one needs to cope with several technical problems.
For instance one has to deal with different authors with identical names,
with scientists who change their name during their career
and with scientists whose name was transcribed into Latin
in several different ways. In general one might think that 
such cases do not occur very often \cite{Ak08}, so they should
not induce statistically significant effects 
for the weighted indices of all other authors, 
but these important problems definitely require further studies.
Some  remarks on selection of bibliometric data
and other practical issues are provided in Appendix B.

We shall now pass to some more general remarks.
Although we tend to agree that the existing bibliometric indices 
can be further developed and improved, we do not claim
there exists a single number capable to quantify scientific
achievement in an unambiguous way.
%
On the other hand, one should not neglect the possibility
of making a wise use of the bibliometric data and various impact factors.
Let us quote however, an opinion of Adler et al. \cite{AET08}, 
"While it is incorrect to say that the impact factor gives no
information about individual papers in a journal, the information
is surprisingly vague and can be dramatically misleading".

Similarly, any bibliometric data should not play the decisive role
during any peer review process.
For instance, working with applications for Advanced Grants 
of European Research Council (ERC) the panels of experts
tried hard to evaluate the quality of the projects
and the scientific achievements of the principal investigator,
not putting too much attention to their scientometric indices.
However, an a posteriori statistical analysis
found a clear correlation \cite{Zy09}
between the final outcome of the 2008 grant competition 
in PE-2 panel and the 
bibliometric benchmark suggested by ERC and used in the proposals: 
the total number of citations of ten papers
chosen by each applicant from his 
list of publications for the last decade.

\medskip

Let us then conclude this article with 
some concrete comments concerning the practical 
usage of scientometric data.
They will be separately  addressed to three groups of readers.

{\bf a) Scientists}. Do well your research, write good papers
and try to publish them in good scientific journals.
Writing your articles cite these papers which should be quoted,
according to the established habits in your field.
Do not care too much about various impact factors and 
indices. Any good scientist
will have sound numbers with respect to any (reasonable)
measure and scientometric indicator.
Do not waste your time and energy for a silly game to
inflate artificially the values of the scientometric indices, 
which might be used to characterize your research output.

{\bf b) Reviewers}. Scientists involved in all
kind of evaluation should make use of their
knowledge of the field and do not treat the bibliometric data
as a definite answer to any question.
During the peer review process all scientometric indicators
should be considered as auxiliary data only.
In a need to characterize the impact of a given article 
published more then three years ago
one should use the number of citations gained 
instead of the impact factor of the journal it was published.
Furthermore, the bibliometric indicators
should always be normalized against the average 
computed for scientists working in the similar field of science 
and in the corresponding period of time.

{\bf c) Managers of science}.
Scientific activity has multiple goals,
so try to avoid harsh consequences of the projection
of a multidimensional system onto a single axis.
Do not hope therefore for a unique scientometric indicator,
which could be widely used as a universal evaluation tool.
Each bibliometric index has certain advantages and some drawbacks,
but using several of them in parallel reduces the risk
of manipulating the data. Support versatile usage of scientometry,
in which the researcher under evaluation takes active part.
For instance, consider the benchmarks used by applicants for the ERC grants:
Any senior researcher selects his ten best papers published in
the last decade and provides the number each of them was cited.
A junior scientist has to choose his best five papers 
published during the recent five years.

\bigskip
To summarize, it is not fair to say that the
bibliometric data carry no valuable information whatsoever.
However, it is not as simple to decode from them a piece of
relevant information, as it may look like at a first glance.
Thus we would not to discourage from usage of scientometric data,
provided they are used in a wise and reasonable way.

\bigskip

{\bf Note added.} After the first version of this work was completed
a new paper by Radicchi et al. was posted in the web 
and later published \cite{RFMV09}.  The authors of this article put
forward a similar idea to apply the PageRank algorithm
to the citation graph,
in which each vertex represents an individual author.
Working with the set of data composed of the collection of
the Physical Review journals published between 1893 and 2006
they concluded that the
numerical values of the weighted indicator obtained in this way
for several physicists  correlates well with their
scientific achievements measured by some
of the main prizes in physics, which include
Nobel prize, Boltzmann medal, Wolf prize, 
Dirac medal and Planck medal.

\medskip 

It is a pleasure to thank P.~Bia{\l}as, W.~Burkot, G.~Hara{\'n}czyk, 
M.~Ku{\'s} and W.~S{\l}omczy{\'n}ski for helpful discussions
and C.~M.~Bender for fruitful correspondence.
This work was performed during the author's spare time
and was not supported by any funding agency.

\appendix

\section{Some exemplary graph matrices and their leading eigenvectors}

In this appendix we provide examples of some simple matrices
and analyze properties of their leading eigenvector.
Although a matrix of a small size $N$ directly represents
only a small citation graph which describes a small group of $N$ scientists, 
it can be also applied to model a huge graph with a sub--graph structure:
each vertex may represent a given field or subfield of science.
Therefore studying even such oversimplified cases
can be helpful in understanding the properties 
of the connectivity matrix of a citation graph and its leading eigenvector.

Let us start with the simplest case of $N=2$,
\begin{equation}
C_2=\left[
\begin{array}
[c]{cc}%
0 & a\\
b & 0
\end{array} 
\right]  \ ,
\quad  \quad
x=\left[
\begin{array}
[c]{c}
\sqrt{a} \\
\sqrt{b} 
\end{array} 
\right]  \ .
\end{equation}
The leading eigenvalue reads
$\lambda=\sqrt{ab}$,
and  in this case the
 weights $x_i$ given be the corresponding eigenvector are proportional to the square root
of the flow between the vertexes. Obviously this is not longer the case
for larger graphs,
\begin{equation}
C_3=\left[
\begin{array}
[c]{ccc}%
0 & a & 0\\
0 &  0  & b \\
c &  0  & 0 
\end{array} 
\right]  \ ,
\quad  \quad
x=\left[
\begin{array}
[c]{c}
a^{2/3}b^{1/3}\\
b^{2/3}c^{1/3}\\
a^{1/3}c^{2/3}
\end{array}
\right]  \ .
\end{equation}
The following numerical example 
shows that the weights given
by the leading eigenvector grow slower than linearly
with the average entry in each row,
\begin{equation}
C_4=\left[
 \begin{array}
[c]{cccc}%
0 & 1 & 1 & 1\\
2 & 0 & 2 & 2\\
3 & 3 & 0 & 3\\
4 & 4 & 4 & 0
\end{array}
\right]  \ ,
\quad  \quad
x \approx \left[
\begin{array}
[c]{c}%
0.3223\\
0.5738\\
0.7755\\
0.9409
\end{array}
\right]  \ .
\end{equation}

Consider now some other numerical  examples of size four
\begin{equation}
C=\left[
 \begin{array}
[c]{cccc}%
0 & 4 & 0 & 0\\
6 & 0 & 2 & 1\\
0 & 0 & 0 & 0\\
0 & 0 & 3 & 0
\end{array}
\right]  \ ,
\quad  \quad
x \approx \left[
\begin{array}
[c]{c}%
0.6325\\
0.7746\\
0\\
0
\end{array}
\right]  \ .
\end{equation}

Observe that quotations by authors, the papers of which were never
cited do not contribute at all to the 
weighting index! 
\begin{equation}
C=\left[
 \begin{array}
[c]{cccc}%
0 & 6 & 6 & 6\\
2 & 0 & 1 & 1\\
0 & 1 & 0 & 1\\
1 & 0 & 1 & 0
\end{array}
\right]  \ ,
\quad  \quad
x \approx \left[
\begin{array}
[c]{c}%
0.8975\\
0.4228\\
0.1247\\
0.2036
\end{array}
\right]  \ .
\end{equation}
Similarly, quotation by a
junior scientist, the papers of which received 
a little attention of the scientific community,
are much less important than a citation
by an accomplished author.
This is seen by comparing the third and the fourth component 
of the eigenvector of the above citation matrix, in which the first
two rows represent a renowed researcher and a less experienced 
author, respectively.

It is illustrative to analyze the case
of two weakly connected subgraphs,
represented below by the first and the second pair
of nodes. If the coupling between the subgraphs is symmetric,
$C_{2,4}=C_{4,2}$
the leading eigenvector lives in both subspaces,
\begin{equation}
C=\left[
 \begin{array}
[c]{cccc}%
0 & 4 & 0 & 0\\
6 & 0 & 0 & 1\\
0 & 0 & 0 & 4\\
0 & 1 & 6 & 0
\end{array}
\right]  \ ,
\quad  \quad
x \approx \left[
\begin{array}
[c]{c}%
0.4197\\
0.5691\\
0.4197\\
0.5691
\end{array}
\right]  \ ,
\end{equation}
However, if there is more fluxes between both subgraphs start to differentiate,
the weight of the leading vector moves toward the distinguished subsystem,
\begin{equation}
C=\left[
 \begin{array}
[c]{cccc}%
0 & 4 & 0 & 0\\
6 & 0 & 0 & 1\\
0 & 0 & 0 & 4\\
0 & 0.5 & 6 & 0
\end{array}
\right]  \ ,
\quad  \quad
x \approx \left[
\begin{array}
[c]{c}%
0.4939\\
0.6502\\
0.3493\\
0.4597
\end{array}
\right]  \ ,
\end{equation}
\begin{equation}
C=\left[
 \begin{array}
[c]{cccc}%
0 & 4 & 0 & 0\\
6 & 0 & 0 & 1\\
0 & 0 & 0 & 4\\
0 & 0.1 & 6 & 0
\end{array}
\right]  \ ,
\quad  \quad
x \approx \left[
\begin{array}
[c]{c}%
0.5913\\
0.7480\\
0.1870\\
0.2365
\end{array}
\right]  \ .
\end{equation}

If two graphs are not connected, the leading eigenvalue is degenerated
and one finds a corresponding
eigenvector localized
exclusively in the more populated subspace, 
\begin{equation}
C=\left[
 \begin{array}
[c]{cccc}%
0 & 4 & 0 & 0\\
6 & 0 & 0 & 0\\
0 & 0 & 0 & 1\\
0 & 0 & 3 & 0
\end{array}
\right]  \ ,
\quad  \quad
x \approx  \left[
\begin{array}
[c]{c}%
0.6325\\
0.7746\\
0\\
0
\end{array}
\right]  \  .
\end{equation}
To lift such a degeneracy one may 
modify the analyzed matrix $C$
by forming its convex combination 
with the flat matrix $S$
such that $S_{ij}=1/N$. 
In this way one assures \cite{BP98}
that the leading eigenvector of $C(p)=(1-p)C+pS$
can be obtained by iterating sufficiently long 
the flat vector with all entries equal, $w_i=1/N$, 
by the matrix $C(p)$.


\section{Practical remarks  on evaluating the weighting vector}

\subsection{Selection of the data}

The key issue by constructing the citation graph is an access to a reliable
data base containing the scientific literature. For instance one may rely
on the data provided by the ISI Web of Science, although some experts
claim that it is biased toward the scientific journals published in English only
and it does not cover uniformly the entire literature.
Alternatively one may chose to use some publicly open web search engines,
like Google Scholar.  In this case it is believed that Google does not cover systematically
earlier scientific literature. Furthermore it is not clear how to set simple criteria,
which web documents should be taken into account. On one hand one might 
restrict the attention to the papers published by a scientific journal,
which is first found in an earlier compiled list of all sources accepted.
On the other hand, due to popularity of various web archives and preprint depositories
(like arxiv.org) one might also accept formally unpublished preprints posted there.
In such a case a special care has to be taken in order
to avoid double counting the same article, first deposited in an archive, and later
published in a journal, often under a slightly changed title.

\subsection{Different fields of science}

  As  illustrated with some simple matrix examples, if two fields of science
  are not coupled by any cross-citations, the leading vector describes
  only scientists working in the larger field. Similarly, if two fields of science 
 are coupled only weakly by a few cross-citations, the leading eigenvector 
 tends to be localized in the subgraph with more scientists, papers and citations,
 so the weighting factors handicap researchers working in a less popular subfield.
The  splitting of the entire graph into subgraphs can be defined in an objective way 
by applying the recent method of Newman \cite{Ne06}
to find community structure in the citation graph.
Since it is well known that the citation patterns depend on the  branch of
science  \cite{BCKM06,AWBB09},
one should rather analyze two subgraphs separately,
or renormalize the leading eigenvector separately for a given subfield.
This is consistent with a rather general  'rule of thumb':
the  bibliometric data should be normalized against the average 
computed for scientists working in the similar field of science 
in the corresponding window of time \cite{RFC08}.

\subsection{Degeneracy in names} 

It might not be easy to distinguish papers written by various 
scientist, who  publish under the very same name \cite{Ak08}.
In principle one may try to distinguish them by 
the scientific discipline, the affiliations and the time window of their 
publishing activity, but is its unlikely to expect that 
the success rate will tend to unity.
On the other hand, it is reasonable to conjecture that
not distinguishing between the scientists with the same name
will not impact much the weighting indices
of all other researchers in the graph, 
as the weights of the links will be taken as  the average.

\subsection{Period of the scientific activity} 

   It would be unwise to compare weighting indices of two researchers 
    in very different age or living in different times.
    The number of universities, scientists, journals,
    papers and citations keeps growing fast. Hence
    one should expect that a comparison of two  scientists with equally 
    valuable accomplishments, whose scientific contributions are already forgotten 
    (and their papers are not quoted any more),
    would reveal that the scientist active more
   recently is characterized by a larger weighting factor.  

\subsection{Uniqueness of the leading eigenvector of the citation matrix}


A matrix $C$ is called {\sl reducible} if it can be transformed by a permutation $P$
into matrix with a zero block below the diagonal, $C'=PCP^T=
\left[  \begin{array} [c]{cc}%
D_1 & Z \\
0 & D_2  \end{array}  \right] $,
 where $D_1$ and $D_2$ are square matrices.
In the opposite case the matrix is called {\sl irreducible}.
The  Frobenius-Perron theorem implies that for any irreducible non-negative matrix
$C$ its {\sl spectral gap} is positive, $\gamma:=z_1-|z_2| >0$,
so the real eigenvector ${\vec x}$ corresponding to the leading eigenvalue $z_1$ is unique.
The size of the spectral gap governs the speed of the convergence of 
any initial vector iterated several times by $C$ 
to the invariant state ${\vec x}=C{\vec x}$.

The initial citation matrix $C$ analyzed in this paper in principle could reducible,
but due to numerous cross-citations between various 
researchers and subfields this possibility seems to be unlikely.
Furthermore, the auxiliary $(N+1)$-th node of the graph 
representing all scientists outside the ensemble
under investigation introduces additional connectivity
and hence increases (on average) the spectral gap.

The size of the spectral gap for the graph matrix describing entire
scientific literature has to be determined in a numerical experiment. 
If the gap occurs to be too small to ensure convergence time realistic 
for practical implementations, 
one may always introduce a suitable modification of the citation matrix $C$.
For instance, following the original idea of Page Rank \cite{BP98},
one could mix $C$ with the flat matrix $S$ such that $S_{ij}=1/N$ --
see also \cite{LM05,BL06}.


\begin{thebibliography}{99}


\bibitem{AR09} D. W. Aksnes and A Rip,
Researchers' perceptions of citations,
{\sl Research Policy}   {\bf  38},   895-905  (2009).  

\bibitem{AET08} R. Adler, J. Ewing, P. Taylor,
   {\sl Citation Statistics},
a report IMU-ICIAM-IMS, November 2008,

\bibitem{LM09} F. Lalo{\"e} and R. Mosseri,
  Not even right, not even wrong,
{\sl Europhysics News}, {\bf 40}/5 27-29 (2009).

\bibitem{gar1} E. Garfield,  {\sl Citation Indexing}
  (Wiley, New York) (1979).

\bibitem{gar2} E. Garfield,  The impact factor,
 {\sl Current Contents} {\bf 29}  20 June 1994.


\bibitem{FG08} M.E. Falafas and V.G. Alexiou,
 The top-ten in journal impact factor manipulations,
{\sl Arch. Immunol. Theor Exp.} {\bf 56}, 223-226 (2008).

\bibitem{Pl94} R. Plomp, 
The highly cited papers of professors as an indicator
of a research group's scientific performance,
 {\sl Scientometrics} {\bf 29}, 377 (1994).

\bibitem{Hi05} J.E. Hirsch,  An index to quantify an individual's scientific
research output,
      PNAS {\bf 102} 16569 (2005).


\bibitem{BP98} S. Brin and L. Page,
The anatomy of a large--scale hypertextual Web search engine,
     {\sl Comp. Networks ISDN Systems} {\bf 30} 107 (1998).

\bibitem{CXMR07} P. Chen, H. Xie, S. Maslov and S. Redner,
Finding scientific gems with Google Pare Rank algorithm, 
{\sl J.Informetrics} {\bf 1}, 8-15 (2007).

\bibitem{MGZ08} N. Ma, J. Guan and Y. Zhao,
Bringing Page Rank to the citation analysis,
{\sl Information Processing \& Management} {\bf 44}, 800 (2008).

\bibitem{Ni08} M. A. Nielsen,
 Lectures on the Google Technology Stack, December 2008, 
 see  http://michaelnielsen.org

\bibitem{Be07} C. Bergstrom, 
Eigenfactor: Measuring the value and the prestige of scholarly journals,
C\&RL News, {\bf 68}, 5 (2007).

\bibitem{AJM04} Y. An, J. Janssen and E. E. Milios,
   Characterizing and mining the citation graph
  of the computer science literature,
   {\sl Knowledge \& Information Systems} {\bf 6}, 664 (2004).

\bibitem{MO79} A. W. Marshall and I. Olkin,
   {\it The Theory of Memorization and Its Applications}
  (Academic Press, New York, 1979)

\bibitem{Ber05:1} D.~S. Bernstein,
{\it Matrix Mathematics},
Princeton University Press, Princeton  (2005).

\bibitem{Ak03} D.W. Aksens,
 A marco-study of self--citations  
{\sl Scientometrics} {\bf 56}, 235 (2003).

\bibitem{GDTS06} W. Gl{\"a}nzel, K. Debackere, B. Thijs, and A. Schubert,
A concise review on the role of author self-citations
 in information science, bibliometrics and science policy,
{\sl Scientometrics} {\bf 67}, 263-277 (2006)


\bibitem{Sch07} M. Schreiber,
 A case study of the Hirsch index for $26$
non-prominent physicists,
{\sl Ann. Physik} {\bf 16},  640-652   (2007).


\bibitem{Fr07}  T. F. Frandsen,
Journal self-citations - Analysing the JIF mechanism,
{\sl J. Informetrics} {\bf 1}, 47-58  (2007)

\bibitem{FA07} J. H.  Fowler and D. W. Aksnes,
 Does self-citation pay?
{\sl Scientometrics} {\bf 72}, 427-437 (2007).


\bibitem{Se97} P.O. Seglen,
 Why the impact factor should not be used for evaluating research
{\sl BMJ} {\bf 324}, 497 (1997).

\bibitem{RFC08}  F. Radicchi, S. Fortunato and C. Castellano,
    Universality in citation distribution:
     towards an objective measure of scientific impact,
{\sl Proc. Natl. Acad. Sci. USA} {\bf 105}, 17268-1727 (2008).

\bibitem{Gl06} W. Gl{\"a}nzel,
On $h$--index. A mathematical approach to a new measure of publication activity
and citation impact,
{\sl Scientometrics} {\bf 67}, 315 (2006).

\bibitem{Hi07} J.E. Hirsch,
 Does the $h$ index have predictive power?
    {\sl PNAS} {\bf 104} 19193 (2007).


\bibitem{Wo08} G. J. Woeginger,
An axiomatic characterization of the Hirsch index,
{\sl Math. Soc. Sciences} {\bf 56}, 224 (2008).

\bibitem{MM08} J.F. Molinari and A. Molinari,
A new methodology for rating scientific institutions,,
{\sl Scientometrics} {\bf 75}, 163 (2008).

\bibitem{Sc07} A. Schubert,
Successive $h$-indices,
{\sl Scientometrics} {\bf 70}, 183 (2007).

\bibitem{BCKM06} P.D. Batista, M.G. Campiteli, O.Kinouchi, and A.S. Martinez,
Is it possible to compare researchers with different scientific interest?
{\sl Scientometrics} {\bf 68}, 179 (2006).

\bibitem{AWBB09}  B. M. Althouse, J. D. West, T. C. Bergstrom, and C. T. Bergstrom,
 Differences in impact factor across fields and over time,
{\sl J. Am. Soc. Inf. Sci. Technol.} {\bf 60} 27-34  (2009).

\bibitem{Re98} S. Redner,
 How popular is your paper? An empirical study of the citation distribution,
{\sl  Eur. Phys. J. } {\bf   B 4}, 131 (1998).

\bibitem{LLJ03} S. Lehmann, B. E. Lautrup, A. D. Jackson, 
Citation Networks in High Energy Physics,
{\sl  Phys. Rev.} {\bf  E 68}, 026113 (2003).

\bibitem{Ba06} M. G. Banks,
   An extension of the Hirsch index: Indexing scientific topics and compounds,
{\sl Scientometrics} {\bf 69}, 161 (2006).

\bibitem{Ak08} D.W. Aksens, 
When different persons have an identical author name. How frequent are homonyms?
 {\sl JASIST} {\bf 59},  838-841 (2008).

\bibitem{Zy09} K. {\.Z}yczkowski,
  How to get an ERC grant? 
{\sl Europhysics News}, {\bf 40}/5, 27-29 (2009).

\bibitem{RFMV09}  F. Radicchi, S. Fortunato,
 B. Markines  and A. Vespignani,
  Diffusion of scientific credits
 and the ranking of scientists,
{\sl Phys. Rev.} {\bf E 80}, 056103 (2009).

\bibitem{Ne06} M. E. J. Newman,
Finding community structure in networks using the eigenvectors of matrices,
{\sl Phys. Rev.} {\bf E 74}, 036104 (2006).

\bibitem{LM05} A.N. Langville and C.D. Meyer,
 A survey of eigenvector methods for web information retrieval,
{\sl SIAM Review} {\bf 47} 135 (2005).

\bibitem{BL06} K. Bryan and T. Leise,
The \$ 25,000,000,000 eigenvector:
  the linear algebra behind Google,
SIAM Review {\bf 48} 569 (2006).


\end{thebibliography}
\end{document}